\documentclass[twocolumn,aps,prd,amsmath,amssymb,nofootinbib,nobibnotes,floatfix,superscriptaddress]{revtex4-1}

\usepackage{graphicx}
\usepackage{float}
\usepackage{bm}
\usepackage[hidelinks]{hyperref}
\usepackage{color}

\usepackage[utf8]{inputenc}
\usepackage[T1]{fontenc}
\usepackage{slashed}

\usepackage{enumerate}

\usepackage{mathalfa,amssymb,amsthm}
\usepackage{amsmath}

\newcommand{\ssec}[1]{\section{#1}}
\newcommand{\rn}{Reissner-Nordstr\"om}

\newcommand{\beq}{\begin{equation}}
\newcommand{\eeq}{\end{equation}}
\newcommand{\beqa}{\begin{eqnarray}}
\newcommand{\eeqa}{\end{eqnarray}}

\begin{document}

\title{
    Gravitational collapse in the vicinity of the extremal black hole critical point 
   }
\author{William E.\ East}
\email{weast@perimeterinstitute.ca}
\affiliation{Perimeter Institute for Theoretical Physics, Waterloo, Ontario N2L 2Y5, Canada}

\begin{abstract}
We study the threshold of gravitational collapse in spherically symmetric
spacetimes governed by the Einstein-Maxwell-Vlasov equations. 
We numerically construct solutions describing a collapsing distribution of charged matter
that either forms a charged black hole or eventually disperses.
We first consider a region of parameter space where the solutions at the threshold of black hole formation
are stationary, horizonless shells.
These solutions terminate at a critical point, with their charge-to-mass ratio approaching unity from below,
and the instability timescale diverging.
Beyond the critical point, we find a new region of parameter space
where the threshold solution is an extremal black hole.  
We measure the scaling of the dynamical time period of the near threshold solutions
and discuss how they are connected 
in the two regimes.
If a similar picture to the one found here holds for 
known families of stationary solutions of rotating matter that approach the exterior of an extremal Kerr
spacetime, they could provide a route to forming an extremal spinning black hole. 
\end{abstract}
\maketitle
\ssec{Introduction}%
There is a close analogy between phase transitions and the general relativistic
dynamics of gravitational collapse near the threshold of black hole formation.
In pioneering work, Choptuik~\cite{Choptuik:1992jv} considered different
families of collapsing massless scalar field pulses, showing that, approaching
threshold, these solutions display universality, self-similarity, and a power
law scaling of the black hole mass (above threshold) and the maximum curvature
(below threshold)~\cite{Garfinkle:1998va}. This is similar to the behavior
found in a continuous phase transition at a critical point.  The critical
solution corresponds to a zero-mass, naked singularity, which means that the
cosmic censorship conjecture at best holds with a genericness condition.

Following this, critical phenomena in gravitational collapse have been studied
in a number of settings and matter models (see Ref.~\cite{Gundlach:2025yje} for
a review).  This includes the collapse of collisionless particles governed by
the Vlasov equation~\cite{Rein:1998uf,Olabarrieta:2001wy,Martin-Garcia:2001jtc,Andreasson:2006gx,Akbarian:2014gna,Gundlach:2016noy,Gundlach:2017jty}
(which is the focus of this work), as well as extensions beyond the
spherically symmetric setting originally
studied~\cite{Abrahams:1993wa,Choptuik:2003ac,Baumgarte:2019fai,Ames:2020ssl,Baumgarte:2023tdh,Marouda:2025wue}.
For some matter models and parts of parameter space, the solution at the
threshold of black hole formation is a (not necessarily unique) gravitationally
bound, but horizonless, starlike solution that is unstable to collapse. In
this case, the black hole mass does not become arbitrarily small approaching
threshold, and this is usually referred to as ``type I" in analogy with a
discontinuous phase transition~\cite{Choptuik:1996yg,Brady:1997fj} (while cases
like that originally studied in Ref.~\cite{Choptuik:1992jv} are referred to as
``type II").

Recently, Kehle and Unger~\cite{Kehle:2024vyt} demonstrated a new regime of
critical gravitational collapse where the threshold solution is an extremal
charged black hole.  Building on earlier work studying thin
shell~\cite{1983GReGr..15..403P} or charged null dust~\cite{Ori:1991}
models, they mathematically constructed solutions to the
Einstein-Maxwell-Vlasov equations corresponding to distributions of charged particles
that either eventually disperse when the total charge exceeds the total mass
$Q>M$ (where we use geometric units with $G=c=1$ throughout), form a black
hole when $Q<M$, or form an extremal horizon (with no trapped surfaces) when
$Q=M$.  Though the black hole mass does not become arbitrarily small
approaching threshold (similar to previously studied type I cases described
above, and in contrast to the type II), the threshold solution is continuously
connected to the black hole forming cases (in contrast to the usual type I
starlike threshold solutions).  Such solutions can also be used to construct
counterexamples to Israel's formulation of third law of black hole mechanics
(alongside the charged scalar field construction in Ref.~\cite{Kehle:2022uvc}),
which states that a black hole cannot have its surface gravity reduced to zero
in finite advanced time~\cite{Israel:1986gqz}; here, the black hole's surface
gravity corresponds to temperature.  An open question raised by this work is
what scaling relations, if any, are exhibited by extremal critical collapse near
threshold?  Another is whether this can be extended to the case of an extremal
rotating black hole.

In this work, we also study the threshold of black hole formation for
collapsing shells of charged matter.  We numerically construct solutions of the
Einstein-Maxwell-Vlasov equations where, in one regime, the threshold solutions
are stationary shell solutions, i.e., a discontinuous phase transition.  This
behavior terminates at a critical point as $Q/M$ approaches unity from below and the
compactness of the stationary solutions approaches that of a black hole. Beyond
the critical point, we find behavior similar to that in
Ref.~\cite{Kehle:2024vyt}, where the threshold solution is an extremal black
hole. We also study how the time for a black hole to form, or for the shell to
disperse, scales with how the total charge of the spacetime differs from the
threshold value on either side of the critical point.  This illuminates how
extremal critical collapse is connected to previously studied critical
collapse, and deepens the analogy with phase transitions. It also provides a
possible route to extending extremal critical collapse to the rotating black
hole case, and constructing counterexamples to the rotating black hole version
of the third law. 

\ssec{Model and Methodology\label{sec:methodology}}%
We evolve the Einstein-Maxwell-Vlasov equations governing a distribution of electrically charged, self-gravitating particles. 
We restrict to spherically symmetric spacetimes, and imagine the matter is made up of particles with uniform charge-to-rest-mass ratio
$q_0$ and angular-momentum-magnitude-to-rest-mass ratio $\ell_0>0$
(though the stress-energy tensor has no associated angular momentum 
due to there being no preferred angular direction for the particle velocities).
We evolve the metric in ingoing Eddington-Finkelstein-like coordinates, 
\beq 
ds^2 = -a b^2dt^2+2bdtdr + r^2 d\Omega^2 \ ,
\eeq
where $t$ is advanced time (commonly denoted by $v$ in other references), $r$ is the areal radius, and $d\Omega^2$ is the metric on the two-sphere. 
The Vlasov equation is evolved using particle-in-cell methods, where we approximate
the distribution function using a collection of particles that each follow the Lorentz force equation.
The one nontrivial component of the Faraday tensor
corresponding to the radial electric field can be determined by integrating over the charge
distribution sourced by the particles at each time. 
Likewise, the metric functions $a$ and $b$ can be determined at each time by integrating
matter source terms sourced by the particles. 
We do this numerically in such a way that, for the outer region of the domain where
there are no particles, we obtain exactly (up to floating point roundoff) the \rn{}  
solution that is determined by the specified total mass $M$ and total charge $Q$. 
Since $a=1-2m(r)/r$, where $m$ is the Hawking mass, the presence of an apparent horizon
is indicated by $a=0$.
The explicit equations and the numerical methods 
used to solve them are given in Appendices~\ref{sec:evo} and~\ref{sec:num_methods}.

We consider initial data describing a radially infalling shell of charged particles
that is constructed in three steps: 
(i) we construct a stationary, but unstable solution describing
a shell of charged particles where the self-gravity and charge repulsion balance each other;
(ii) we evolve this solution in \textit{outgoing} Eddington-Finkelstein-like coordinates
(or equivalently, we evolve backwards in advanced time), 
letting numerical truncation error excite a linear instability and cause the shell to disperse;
(iii) we take the data at some time after the shell has
sufficiently dispersed\footnote{We evolve these solutions until $\min{a}=0.85$.}, time-reverse it ($t\rightarrow -t$), and apply a 
perturbation to $q_0$ and (in some cases) $\ell_0$ while keeping the particle positions and radial momentum
fixed (and then resolve for the metric and electric field).
This procedure then results in data describing an infalling spherical shell that can be evolved in
ingoing Eddington-Finkelstein-like coordinates.

The static solutions\footnote{The spacetime is static, though the particles have non-zero velocity.} used in step (i) are similar to those considered
in Ref.~\cite{Andreasson:2009qu}, consisting of shells with a distribution of radial momenta
at each point. In particular, here we choose the distribution function to be linear in energy,
up to some specified maximum energy.
Spatially, the particle distribution has nonzero support only within
some range $R_{\rm in}<r<R_{\rm out}$. 
Inside the shell the electric field is zero and the spacetime is flat,
and outside the shell the spacetime
is the same as \rn{}. 
For definiteness, we restrict to static solutions with $q_0=1.25$ for the main results,
though qualitatively similar results can be obtained for other distribution functions or values of $q_0>1$~\cite{Andreasson:2009qu}. 
(For smaller values $q_0$, we do not expect extremal black hole formation~\cite{Reall:2024njy}.)
With this choice, the static shell solutions we consider form a one parameter family that 
can be labeled either by the ratio of the total charge $Q$ to total mass $M$ of the spacetime, or by $\ell_0/M$.
In particular, we will consider these solutions in the region of parameter space where $Q/M$ approaches
unity from below as $\ell_0$ approaches some value $\ell_c$ from above.
More details on the static solutions and how they are constructed are given in 
Appendix~\ref{sec:static_soln}. 
Details on numerical resolution and convergence are given in Appendix~\ref{sec:conv}.

For the results presented here, we explore the parameter space in the following
way. For $\ell_0>\ell_c$, we consider different corresponding members of the static solution
family in step (i) above, and then, in step (iii), vary $q_0$ to form
a one parameter family crossing the black hole formation threshold.
For $\ell_0<\ell_c$, we always consider the same fixed member of the static solution family 
with $\ell_0$ slightly above $\ell_c$ in step (i), and then, in step (iii),
decrease $\ell_0$ by some fixed amount while again varying $q_0$ to form
a one parameter family crossing the black hole formation threshold.

\ssec{Results}
\begin{figure}[h!]
\begin{center}
\includegraphics[width=\columnwidth,draft=false]{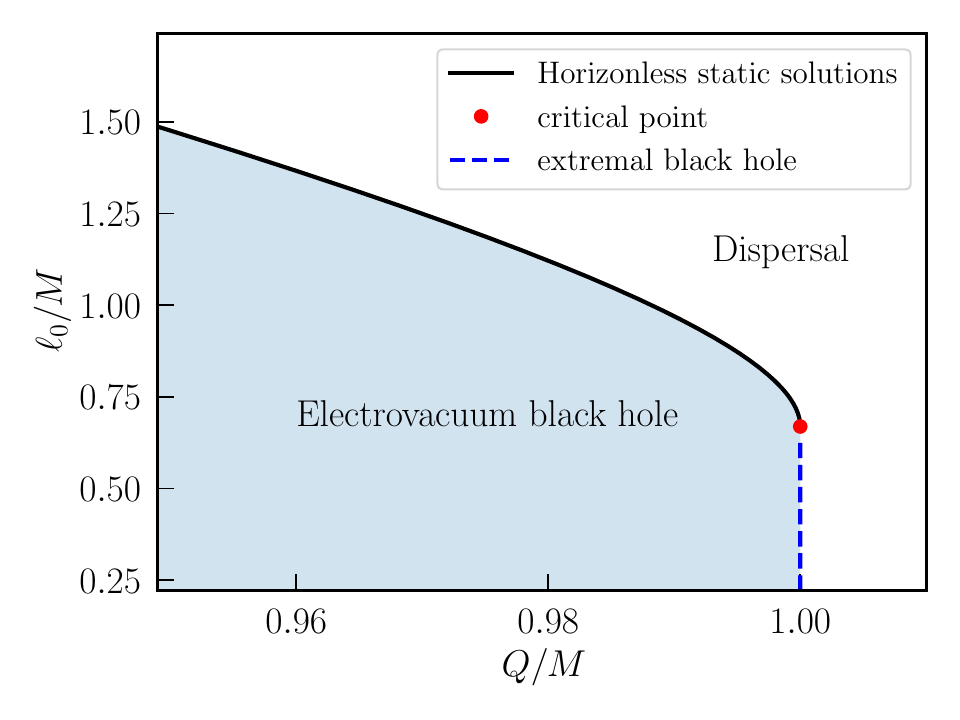}
\includegraphics[width=\columnwidth,draft=false]{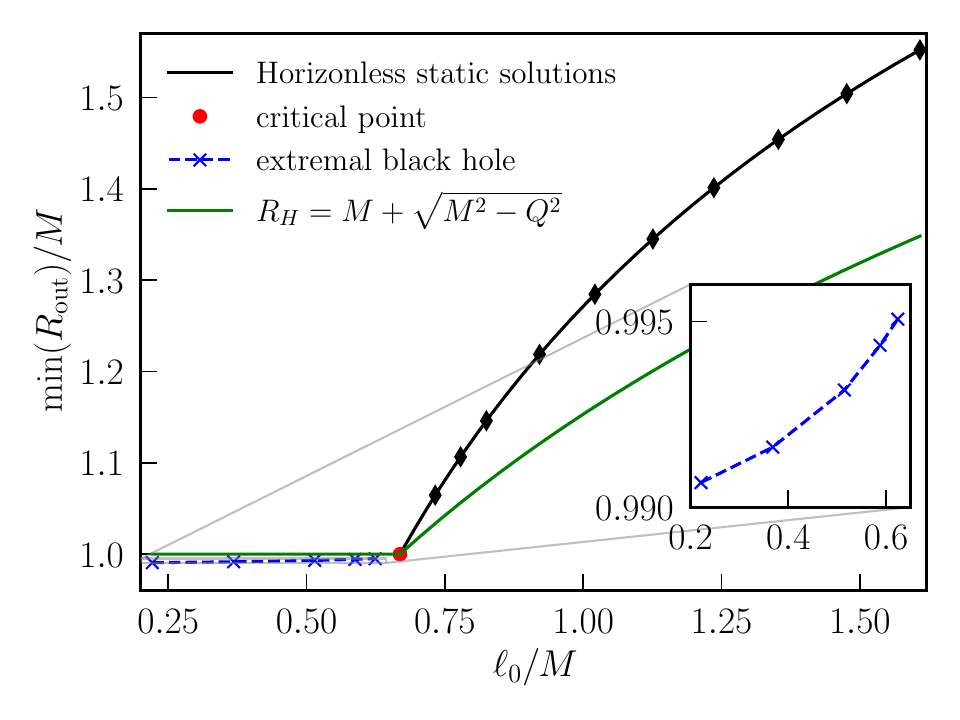}
\end{center}
\caption{
Top: the phase diagram showing the regions of parameter space that result in a black hole (shaded region) versus those that result in
a dispersing charged shell (white region). The $x$ axis is the total charge of the spacetime (and also the charge of the black hole when it forms) and the $y$ axis indicates the particle angular momentum.
For higher values of the particle angular momentum, the threshold solutions at the boundary between the two regions are static shell solutions (solid black line) that terminate at a critical point (red dot) as $Q/M\rightarrow1$. For lower values, the threshold solution is an extremal black hole (dashed blue line). 
Bottom: the outer (areal) radius of the threshold solutions, beyond which the spacetime is electrovacuum, as a function the particle angular momentum. 
For the extremal black hole cases (also shown in the inset), this is the minimum radius the matter reaches inside the horizon. 
The crosses and diamonds show the specific parameters evolved. The green curve indicates the radius of a \rn{} black hole with the same charge as the threshold solution.
\label{fig:phase_diagram}
}
\end{figure}
Within the considered family of initial data describing an infalling shell of charged matter, we
are interested in delineating the boundary between cases where black hole formation 
does and does not occur.
The results we find are summarized by the phase diagram in the top panel of Fig.~\ref{fig:phase_diagram}.
For larger values of the particle angular momentum $\ell_0$, 
the boundary of black hole formation corresponds to the unstable, static charged shell solutions
indicated by the black line. Solutions with slightly greater 
charge or particle angular momentum than the static solutions correspond to shells that fall to some radius and then become outgoing, eventually
dispersing due to the charge repulsion and angular momentum barrier. Solutions with slightly less charge or angular momentum
correspond to infalling shells that form charged black holes, with all the matter falling into the horizon, leaving a \rn{} 
exterior solution.
For decreasing $\ell_0$ approaching some critical value ($\ell_c/M\approx0.67$ for this case), the charge-to-mass ratio of the static threshold solutions represented by the black
curve increases, approaching $Q/M \rightarrow 1$ from below. In this limit, the solutions become more and more compact, with the outer radius $R_{\rm out}$, beyond which the solution is electrovacuum,
approaching $M$ (the radius of an extremal black hole) from above (saturating the bound in Ref.~\cite{Andreasson:2008xw}, as in Ref.~\cite{Andreasson:2009qu}). This is shown in the bottom panel of Fig.~\ref{fig:phase_diagram}.
This limit is singular and
the maximum charge density of the static solutions also blows up as $Q/M \rightarrow 1$ as these solutions terminate at a critical point (indicated in red).

When we further decrease the value of $\ell_0/M$ for the infalling shell solutions
below this critical point, we find that the solutions with $Q>M$ eventually
fully disperse, while for those with $Q<M$, all the particles fall into a black
hole, with the threshold solution always corresponding to an extremal black hole with
$Q=M$. In the bottom panel of Fig.~\ref{fig:phase_diagram}, we show the specific
cases evolved, as indicated by the marked points (crosses and diamonds). 
In particular, for the solutions in the vicinity of the static shell solutions, the highest value of $Q/M$ we consider is $0.9994$ with $R_{\rm out}/M=1.065$ (i.e., the black diamond closest to the critical point). 

It is this static solution (with $\ell_0/M\approx 0.73$) that we leverage,
following the procedure described in the previous section, to construct the initial data
for the evolutions where the threshold solution is an extremal black hole,
by taking the time reversal of the dispersed unstable static solution and decreasing $\ell_0$ while tuning $q_0$.
In the bottom panel of Fig.~\ref{fig:phase_diagram}, we also show the minimum of the outer radius (as a function of time, indicated by the blue crosses) that the matter reaches in the cases
where an extremal black hole forms (technically, we choose $0<1-Q/M<10^{-12}$).
From there, it can be seen that the electrovacuum region extends a small but resolved\footnote{
We estimate that $\min R_{\rm out}$ is accurate to better than $0.1\%$, and that finite resolution actually
leads to an overestimate. See Appendix~\ref{sec:conv}.} amount inside the horizon, $\min{R_{\rm out}}\lesssim 0.995M$, which increases as $\ell_0$
is decreased.

\begin{figure}
\begin{center}
\includegraphics[width=\columnwidth,draft=false]{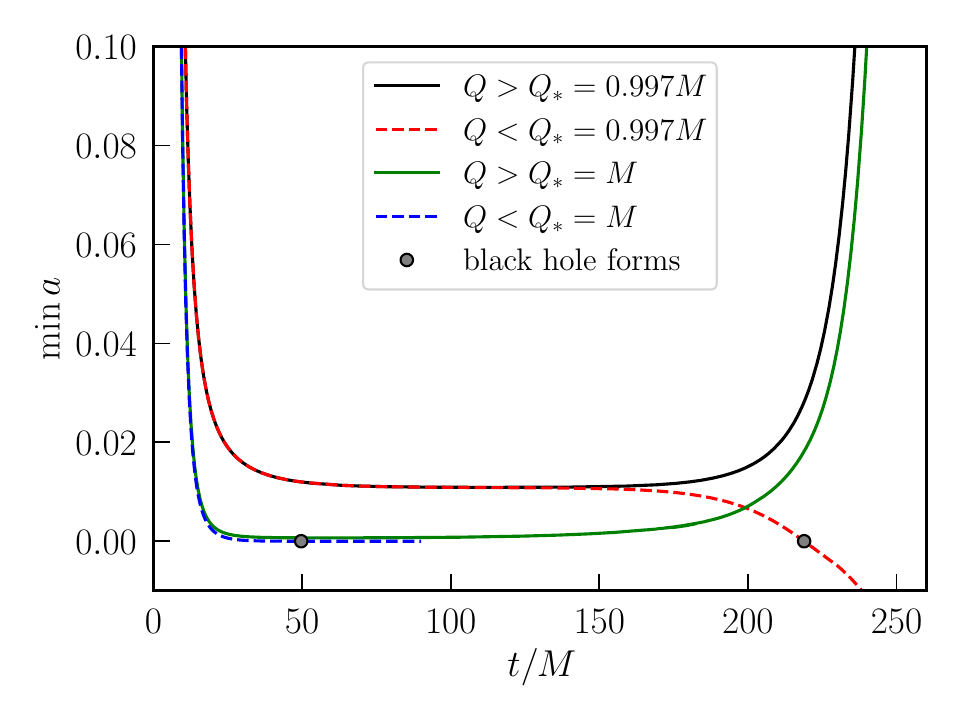}
\includegraphics[width=\columnwidth,draft=false]{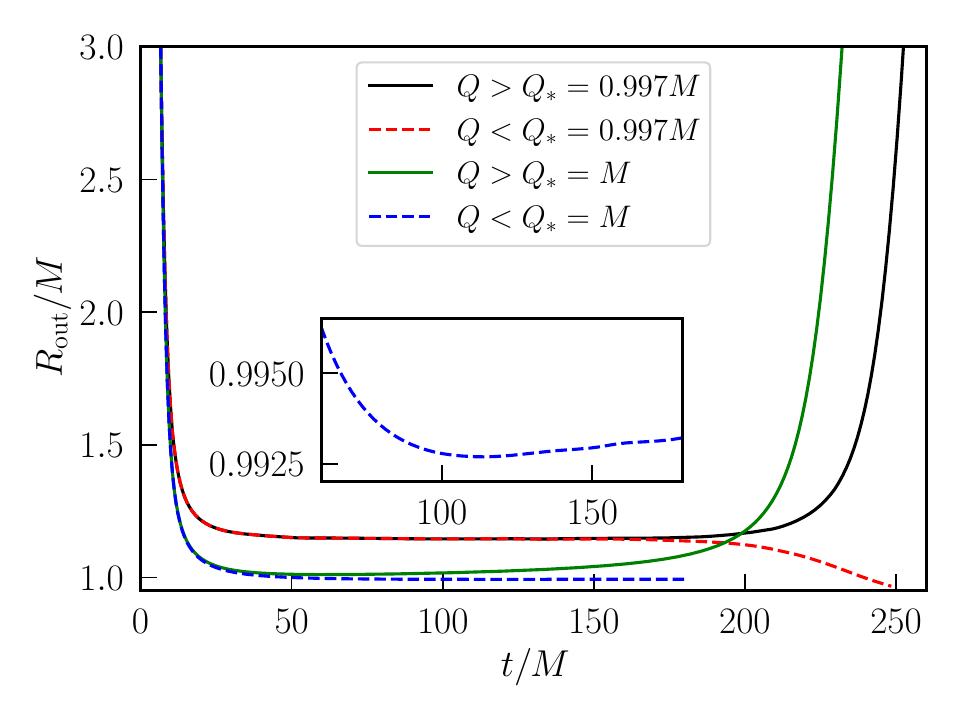}
\end{center}
\caption{
Example evolutions as a function of advanced time just above (solid lines) and below (dashed lines) the threshold for
black hole formation in the two regimes, where the threshold
solution is either a horizonless shell (black and red curves) or an extremal black hole (blue and green curves). 
Top: the minimum value of the metric function $a(r)=1-2m(r)/r$ at each time. 
The circles indicate when a horizon first appears for the dashed lines ($a=0$). 
Bottom: the outer radius of the region containing the particles as a function of time.
The inset shows a zoom-in to illustrate that $R_{\rm out}$ reaches a minimum
slightly below $M$ for the case where $1-Q/M\approx10^{-12}$. 
\label{fig:evolve}
}
\end{figure}

For illustrative purposes,
we will now concentrate on two representative cases, corresponding to $\ell_0\approx 0.825M>\ell_c$ (with threshold charge $Q_*/M\approx 0.997$)
and $\ell_0\approx0.515M<\ell_c$ (with threshold charge $Q_*/M=1$). 
We show the behavior just above and below the threshold for black hole formation for the 
two cases in Fig.~\ref{fig:evolve}.
When the threshold solution is a horizonless static solution (with total charge $Q_*<M$),
a near threshold solution corresponds to a shell that falls inward, spends a long time close to the 
unstable static solution, and then either disperses when $Q>Q_*$, or collapses further inward and 
forms a black hole when $Q<Q_*$ (here $|Q/Q_*-1|\sim 10^{-8}$).
The advanced time that it takes either to disperse or to form a black hole (as indicated by an apparent horizon) becomes longer and longer the closer
one tunes to the threshold.

When the threshold solution is an extremal black hole ($Q_*=M$), the time 
for the shell to fall inward and then disperse when $Q>M$ (we show $Q/M-1\approx10^{-4}$ in Fig.~\ref{fig:evolve}) 
again becomes longer and longer the closer 
one is to threshold. However, for $Q\leq M$ (we show $1-Q/M\approx10^{-12}$ in Fig.~\ref{fig:evolve}), the black hole always forms promptly.
As illustrated in the inset in Fig.~\ref{fig:evolve}, sufficiently close to threshold, $R_{\rm out}$ reaches a minimum value $<M$ before increasing again due to the angular momentum barrier and charge repulsion.
Above threshold, the matter expands outward and presumably approaches a Cauchy horizon in the black hole interior (see Ref.~\cite{Kehle:2024vyt}), though this is difficult
to resolve with our present setup; below threshold, the matter can instead expand outward into the asymptotic region at large radii.

\begin{figure*}
\begin{center}
\includegraphics[width=0.666\columnwidth,draft=false]{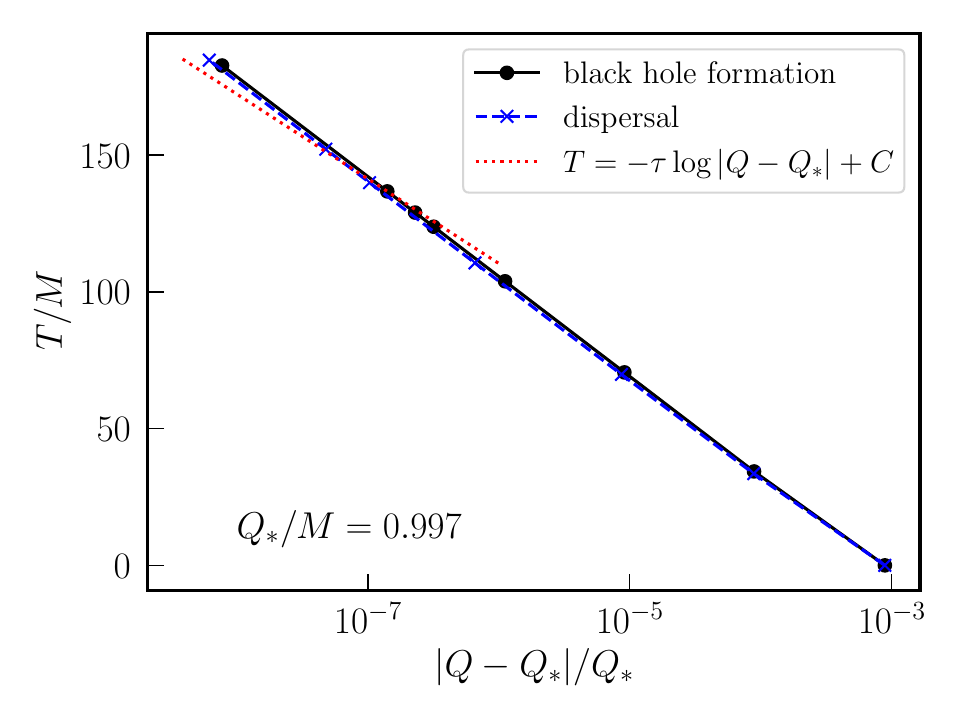}
\includegraphics[width=0.666\columnwidth,draft=false]{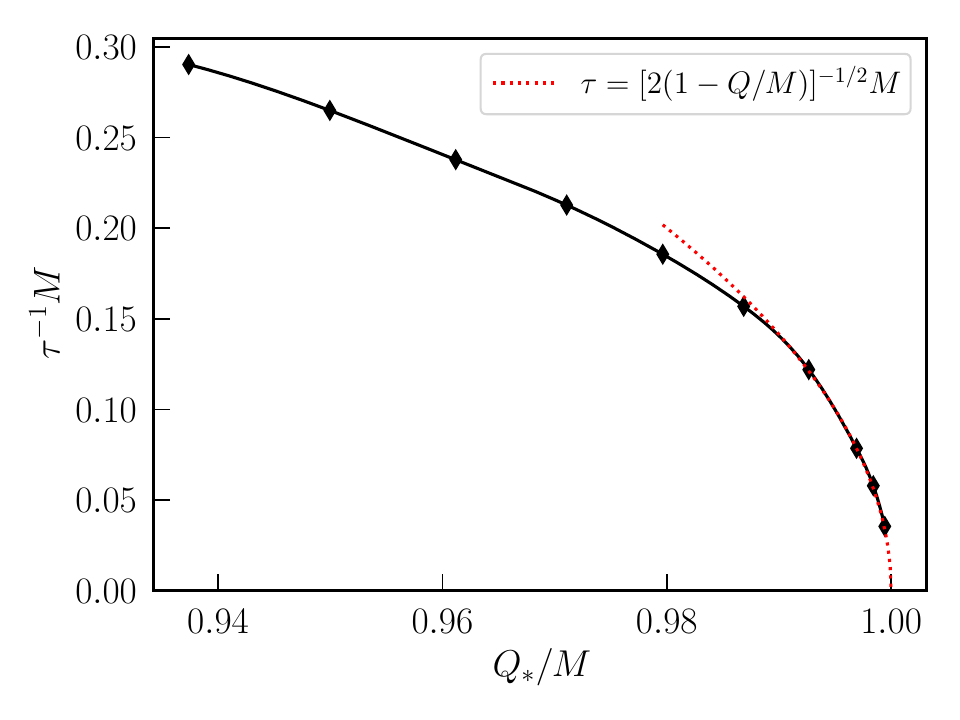}
\includegraphics[width=0.666\columnwidth,draft=false]{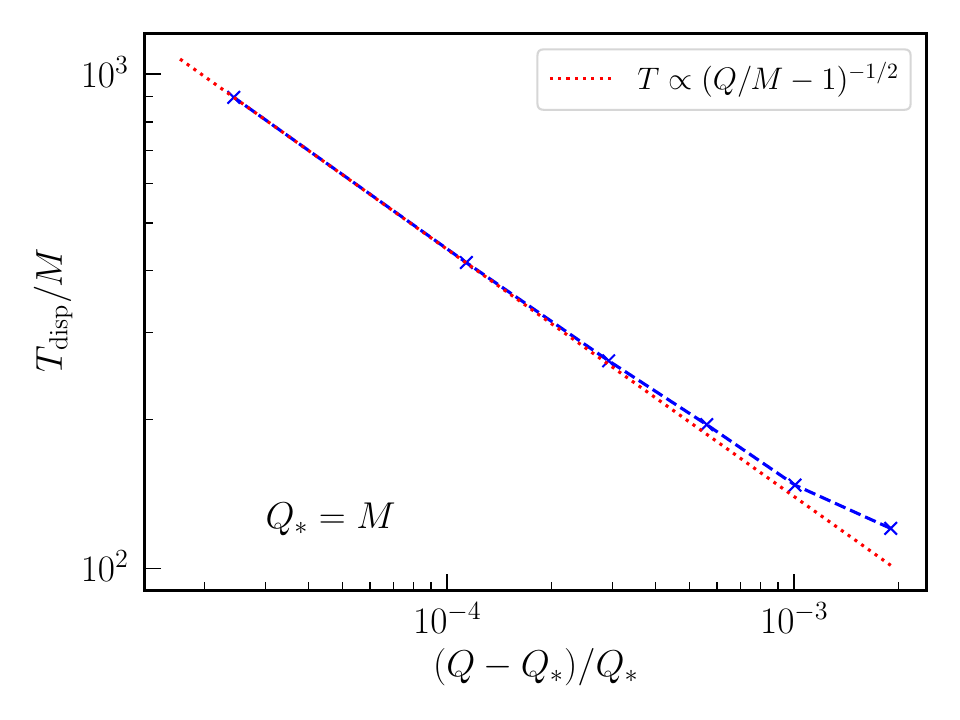}
\end{center}
\caption{
Timescales governing near threshold behavior above (left and center panels) and below (right) the critical particle angular momentum $\ell_c$. 
Left: the time it takes for a black hole apparent horizon to form (black) or for the charged shell to disperse (blue, defined as $R_{\rm out}/M=10$) as a function of the difference in the total charge from its threshold value. 
Time is measured as the difference in the ingoing Eddington-Finkelstein coordinate from when this occurs when $|1-Q/Q_*|=10^{-3}$. 
This approximately matches the dependence expected from measuring the instability timescale of the static threshold solution (dashed red curve).
Center: the inverse instability timescale (i.e., instability rate) for the static charged shell solutions, as a function of their total charge. 
The dotted red line shows an analytic prediction from the surface gravity of a near extremal black hole with the same mass and spin.
Right: the time it takes for the charged shell to disperse (defined as $R_{\rm out}/M=10$) as a function of how
much the total charge is above the threshold value of $Q_*=M$ (for
$\ell_0/M\approx0.52$). This matches well with a $-1/2$ power law scaling
(dotted red line).
\label{fig:shell_timescale}
}
\end{figure*}

We can measure how these different times scale as one approaches the 
critical solution\footnote{As noted above, we form one parameter families crossing threshold by varying $q_0$ while keeping the position and velocities
of the infalling distribution of particles fixed. This predominately changes the total charge $Q$, but also has 
a small effect on $M$. Thus, these families would appear as almost horizontal lines on the top panel of Fig.~\ref{fig:phase_diagram}, 
but with a small negative slope ($\Delta \ell_0/\Delta Q \sim -0.1$ for the cases discussed below).}.
For $Q_*<M$, one expects that both the time for
the black hole to form or the time for the shell to disperse (depending
on whether one is above or below threshold) should 
be given by
\beq
T = -\tau \log |Q-Q_*| + C
\eeq  
near threshold, where $\tau$ is the e-folding time associated with the linear
instability and $C$ is some constant. We indeed find that to be the case, as shown for one example in the left panel 
of Fig.~\ref{fig:shell_timescale}. The center panel in Fig.~\ref{fig:shell_timescale} shows 
how the instability timescale $\tau$ varies 
as one considers more and more compact (but still horizonless) threshold
solutions. Approaching the critical point, the instability timescale
diverges like 
\beq
\tau \approx M[2(1-Q_*/M)]^{-1/2} \approx \kappa^{-1} \ ,
\label{eqn:tauhalf}
\eeq
where $\kappa = (r_+-r_-)/(2r_+)$ with $r_{\pm}=M\pm\sqrt{M^2-Q^2}$ is the surface gravity, 
and the approximation holds to leading order in the extremal limit.

Now considering the case where the threshold solution is an extremal black hole ($Q_*=M$),
we can measure the time for the solution to eventually disperse as $Q$ approaches $M$ from above. 
We show this in the right panel of Fig.~\ref{fig:shell_timescale}.
We find that this time also scales like
\beq
T \sim M(Q/M-1)^{-1/2} \ .
\label{eqn:thalf}
\eeq
In this sense, the exponents governing the dispersal time on either side of the critical point  
match (though the scaling of the time for black hole formation does not).

The above scaling can also be derived by considering the orbits of a charged test particle near the horizon
of a near-extremal \rn{} black hole. When $q_0>\sqrt{\ell_0^2+1}$, there are circular orbits at radii approaching the horizon as $Q/M\rightarrow 1$;
these are unstable with a Lyapunov timescale that has the scaling~\cite{Kan:2021blg}
$\lambda_{RN} \approx \kappa$
in this limit. That is, a small perturbation to the radius of the circular orbit radius grows like $\delta r = \delta r(t=0) e^{\lambda t}$. 
Roughly, we can think of the outer edge of the charged shell as behaving in a similar way to such particle orbits. (Note, though,
that the stationary spherical shell solutions have a distribution of radial velocities.) 
Now letting $Q$ approach $M$ from above, if we consider the advanced time it will take a null geodesic to go from some fixed initial radius $<M$
to some fixed large radius $\gg M$ in near but superextremal \rn{}, which gives a lower bound on the time for a timelike trajectory, we likewise obtain the scaling given in Eq.~\eqref{eqn:thalf}. 

In  \rn, a similar $1/2$ power law scaling also governs the amount by which the area and radius
of a near-extremal horizon differ from their extremal values,
and this was also found to hold for the difference in the final horizon area and location in a broader class of dynamical spherically-symmetric
spacetimes including a perturbing neutral scalar field~\cite{Murata:2013daa,Angelopoulos:2026bez}.  

\ssec{Extending to spinning black holes}
We now briefly discuss how the picture found here for charged black holes might
be extended to spinning black holes. Beginning with the work of
Ref.~\cite{Bardeen:1971eba}, there have been several studies constructing
stationary, axisymmetric solutions of uniformly rotating
dust~\cite{Neugebauer:1995pm}, fluid with
pressure~\cite{Ansorg:2002vh,Fischer:2005uy}, or Vlasov matter~\cite{Ames:2016coj,Ames:2018xqt}
demonstrating families of solutions where the matter velocity becomes ultrarelativistic, 
while the total angular momentum approaches
extremality $J\rightarrow M^2$ from above, and the spacetime approaches the exterior of 
extremal Kerr. In this limit, the matter distribution becomes a thin ring concentrated
near the equatorial plane. 
However, the stability or instability of these solutions has not been explored.
Timelike, circular geodesics in the equator of Kerr are unstable for radii approaching 
the photon ring, suggesting that the ring solutions will be similarly unstable approaching
the critical point. 
For a circular, equatorial null geodesic in Kerr, approaching extremality, the Lyapunov exponent
is given by 
\beq
\lambda /\omega_0 \approx (2M\kappa) \approx 2^{1/2}(1-J/M^2)^{1/2}
\eeq
where we have normalized by the orbital frequency of the null geodesic (e.g., Ref.~\cite{Pretorius:2007jn}).
Thus, we can conjecture that these ring solutions would be subject to an instability
where small perturbations would cause them to disperse, or perhaps partially collapse to a Kerr black hole,
but that the associated instability timescale would become longer and longer as $J\rightarrow M^2$ with the above scaling.
Following a similar procedure to the one used above to extend beyond the critical point might then
be a way to construct solutions describing an infalling ring of rotating matter that forms an extremal Kerr black hole
in finite time.
One possible complication to this picture is that even restricting to axisymmetry there would
be gravitational radiation. (Furthermore, allowing nonaxisymmetric perturbations would open up the possibility
of new instabilities, like the ergoregion instability~\cite{Friedman:1978ygc}.)
 
\ssec{Discussion and Conclusion}%
In this work, we have studied the threshold of the formation of a black hole
with large charge in the Einstein-Maxwell-Vlasov model. We find a phase diagram where
extremal black hole formation emerges as the threshold solution when extending
beyond the critical point marking the end of a discontinuous (type I) transition where the
threshold solution is a stationary charged shell.  In the latter regime, the
lifetime of the intermediate stationary shell solution scales logarithmically 
with the difference of the total charge from its threshold value, approaching
$T\sim \kappa^{-1}\log |Q-Q_*|$, where $\kappa$ is the surface gravity of the
black hole that forms around threshold.  

Beyond the critical point, in the regime where the
threshold solution is an extremal charged black hole, the dispersal time
exhibits a $\sim|1-Q/M|^{-1/2}$ scaling (in advanced time) when a black hole does not form, though
when a black hole does form, it forms promptly. 
In this regime, black hole formation switches on at zero surface gravity
(though at nonzero mass and horizon area), and above or below threshold, the
solutions correspond to charged shells that fall to some minimum radius and
then expand outward. The sharp distinction is just whether or not the matter
eventually disperses to the asymptotic region, or is instead confined behind
the black hole horizon.

The above picture shares some characteristics with canonical examples of phase transitions,
like the liquid-vapor transition, where a discontinuous (type I) phase transition terminates
at a critical pressure and temperature. Approaching the critical point, 
the jump in, e.g., density, across the phase transition goes to zero. At the critical
point the transition is continuous (type II); beyond this there is a supercritical fluid
without a sharp phase distinction.

By leveraging stationary solutions, we have found an alternative route, from
that of Ref.~\cite{Kehle:2024vyt}, to
constructing dynamical solutions
where an extremal black hole forms in finite time.
In future work, it would be particularly interesting to determine if stationary
solutions describing rotating rings of
matter~\cite{Bardeen:1971eba,Neugebauer:1995pm,Ansorg:2002vh,Fischer:2005uy}
could likewise be leveraged to construct dynamical solutions describing critical
collapse with an extremal Kerr black hole as the threshold solution, as well as
counterexamples to the third law of black hole mechanics for rotating black
holes. Based on test particle orbits, we speculate that the scaling will be
the same in the rotating and charged extremal critical collapse cases.  
In Ref.~\cite{Yang:2022okb}, this same scaling was also arrived at by invoking 
quasinormal modes in the near extremal Kerr limit.

\acknowledgments
The author thanks Patrick Brady, Christoph Kehle, and Ryan Unger
 for useful discussions. 
The author acknowledges support
from a Natural Sciences and Engineering Research Council of Canada Discovery Grant and an Ontario Ministry
of Colleges and Universities Early Researcher Award.
Research at Perimeter Institute is
supported in part by the Government of Canada through the Department of
Innovation, Science and Economic Development Canada and by the Province of
Ontario through the Ministry of Colleges and Universities. 
This research was enabled in part by support provided by SciNet
(www.scinethpc.ca) and the Digital
Research Alliance of Canada (www.alliancecan.ca). Simulations were performed on
the Symmetry cluster at Perimeter Institute and the Trillum cluster at the
University of Toronto.

\bibliographystyle{apsrev4-1.bst}
\bibliography{ref}

\appendix

\section{Evolution Equations} 
\label{sec:evo}
We evolve the Einstein-Maxwell-Vlasov equations governing a distribution of electrically charged
and self-gravitating particles with charge $\mathfrak{e}$ and rest mass $\mathfrak{m}$.
The Vlasov equation describing the evolution of the distribution function $f(t,x^i,p^i)$ for the particles is given by
\beq
\frac{\partial  f}{\partial t} + \frac{p^i}{p^t}\frac{\partial f}{\partial x^i} -\frac{1}{p^t}\left ( \Gamma^i_{ab} p^a p^b + \mathfrak{e} p^aF_a{}^i\right ) \frac{\partial f}{\partial p^i}=0  \ , 
\eeq
where $F_{ab}$ is the Faraday tensor, $\Gamma^a_{bc}$ is the Christoffel symbol, and the index $i$ runs
over spatial indices.
The four momentum of the particle $p^a=(p^t,p^i)$ satisfies the normalization condition $p_ap^a=-\mathfrak{m}^2$.
The contribution to the stress-energy tensor is then given by 
\beq 
T_{ab} = -\int p_a p_b f \sqrt{-g} \frac{d^3 p}{p_t} \ , 
\eeq
while the current sourced by this matter is
\beq
J^a = - \mathfrak{e} \int p^a f \sqrt{-g} \frac{d^3 p}{p_t} .
\eeq
The Maxwell equations are 
\beq
\nabla_b F^{ab}=4 \pi J^a \text{ and } \nabla_{[a} F_{bc]}=0 ,
\eeq
and the electromagnetic field contributes to the stress-energy tensor as
\beq
T^{\rm EM}_{ab} = \frac{1}{4\pi} \left( F_a{}^cF_{bc}-\frac{1}{4}g_{ab}F_{cd}F^{cd}\right)
\ .
\eeq
The Einstein equations (in terms of the Einstein tensor) are then
\beq
G_{ab}=8\pi(T_{ab}+T^{\rm EM}_{ab}).
\eeq

We assume a spherically symmetric spacetime and use Eddington-Finkelstein-like
coordinates (e.g., Ref.~\cite{Andreasson:2009ua}) where the line element is given by
\beq
ds^2 = -a b^2dt^2+2bdtdr + r^2 d\Omega^2 \ .
\eeq
Here $r$ is the areal radius.
(In other references, it is common to use $v$ in place of $t$ when using ingoing coordinates,
and $u$ when using outgoing coordinates.)
There is one independent component to the Faraday tensor
which can be found from  
\beq
\partial_r q = 4 \pi r^2 \rho_q \ , 
\label{eqn:q}
\eeq
where $q:= |b| r^2 F^{tr}$ is the total charge enclosed within a given radius,
 and $\rho_q:=bJ^t$ is the charge density.
The metric functions are determined by
\beq 
\partial_r \log |b| = 4 \pi r T_{rr}
\label{eqn:b}
\eeq
and 
\beq 
a=1-\frac{2m}{r} ,
\eeq
where 
\beq 
\frac{1}{b}\partial_r \left(mb \right ) = 2\pi r^2(T_{rr}+S)+\frac{q^2}{2r^2} .
\label{eqn:m}
\eeq
Here we have defined
\beq 
S = \frac{-2}{b}T_{tr}-aT_{rr} \ .
\eeq
For $r=0$, $q=0$, $m=0$, and $a=1$. 
As $r\rightarrow \infty$, $q\rightarrow Q$, $m \rightarrow M$, $a\rightarrow 1$ and $b\rightarrow \pm 1$ (for ingoing or outgoing, respectively). 

In these co\"ordinates, the Lorentz force equation for a particle with rest mass $\mathfrak m$, charge $\mathfrak e$, and angular momentum squared $\mathfrak l^2=p_\Omega^2$ takes the form
\begin{subequations}
\label{eqn:rpr}
\begin{align}
\label{eqn:r}
\frac{dr}{dt} =& \frac{1}{2}ab-\frac{b}{2p_r^2}(\mathfrak{m}^2+\mathfrak{l}^2/r^2)
   ,\\
\frac{d p_r}{dt} =& -\frac{p_r}{2}\left (a\partial_r b+ b\partial_r a \right)
-\frac{\partial_r b}{2p_r}(\mathfrak{m}^2+\mathfrak{l}^2/r^2) \nonumber
\\&+\frac{b \mathfrak{l}^2}{p_r r^3}+\frac{ \mathfrak{e}|b|q}{r^2}
\label{eqn:pr}
   .
\end{align}
\end{subequations}
In all cases here, we consider a distribution of particles with fixed charge to rest mass ratio $q_0=\mathfrak e/\mathfrak m$ and angular momentum magnitude to rest mass ratio $\ell=\mathfrak l/\mathfrak m$. 

\section{Numerical methods}
\label{sec:num_methods}
We numerically evolve the Einstein-Maxwell-Vlasov equations using particle-in-cell methods\footnote{
The code used to construct all the numerical solutions in this paper is publicly available at 
\url{https://bitbucket.org/weast/sph_ein_max_pic/}.
}.
We use $N$ particles to approximate the distribution function, together with a numerical grid
that is regularly spaced in the compactified radial coordinate $\bar{r}:=r/(1+r)$, where the domain
is given by $\bar{r}\in[0,1]$; that is, the grid extends to spatial infinity.
For each particle, we evolve the radial position $r$ and radial momentum $p_r$ by integrating Eq.~\eqref{eqn:rpr}
using fourth-order Runge-Kutta. 
The time component of the particle four momentum is determined
through the normalization condition $p_ap^a=-\mathfrak{m}^2$.
The fields $a$, $b$, their radial derivatives, and $q$
are interpolated to the particle positions using linear interpolation. 

To determine $q$, $b$, and $a$ at the grid points, we numerically integrate Eqs.~\eqref{eqn:q},~\eqref{eqn:b}, and~\eqref{eqn:m} using  
extended Newton-Cotes, a fourth order accurate integration scheme. The particle source terms
are calculated by having each particle contribute a fraction to the two vertices it lies between, weighted
linearly by its distance to each. 
Explicitly, the sources are given by summing over all particles as
\begin{subequations}
\begin{align}
\rho_q(\bar{r}) &= \sum_{n=1}^{N} w(\bar{r},\bar{r}_n)\frac{\mathfrak{e}}{\Delta V(r_n)}\ ,\\
T_{rr}(\bar{r}) &= \sum_{n=1}^{N} w(\bar{r},\bar{r}_n)\frac{(p_r)_n}{\Delta V(r_n)}\ ,\\
S(\bar{r}) &= \sum_{n=1}^{N} w(\bar{r},\bar{r}_n) \frac{\mathfrak{m}^2+\mathfrak{l}^2/r_n^2}{\Delta V(r_n) (p_r)_n}\ .
\end{align}
\end{subequations}
Here 
$\Delta V=(4\pi/3)(r_{i+1}^3-r_{i}^3)$ is 
the coordinate volume of the grid cell the particle lies in ($r_{i}<r_n<r_{i+1}$),
and 
\beq
w(\bar{r},\bar{r}_p)=
\begin{cases} 
|1-(\bar{r}-\bar{r}_p)/\Delta \bar{r}| & \text{if } |\bar{r}-\bar{r}_p|< \Delta \bar{r} \\
0 & \text{otherwise}
\end{cases}
\eeq
 is a weight function that is linear in the compactified coordinate
distance from the particle position $\bar{r}_p$ with compact support over
twice the width of one grid cell $\Delta \bar{r}$. 
Radial derivatives of $a$ and $b$ are solved for directly using Eqs.~\eqref{eqn:b} and~\eqref{eqn:m}.

Equation~\eqref{eqn:b} can be integrated from spatial infinity to give
\beq 
b = \pm \exp( -4 \pi \int_r^{\infty} r' T_{rr} dr') \ , 
\eeq
where the plus (minus) sign refers to ingoing (outgoing) co\"ordinates. 
When using outgoing co\"ordinates, we integrate Eqs.~\eqref{eqn:q} and~\eqref{eqn:m} beginning from $r=0$, where $q=0$ and $a=1$ ($m=0$). 

When using ingoing co\"ordinates, where a black hole interior may be included in the domain, we fix the total charge $Q$ and total spacetime mass $M$ at $t=0$, and integrate Eqs.~\eqref{eqn:q} and~\eqref{eqn:m} from spatial infinity. Explicitly, we solve
\beq
q = Q-\int_r^\infty \rho_q dr'
\eeq  
and
\beq
m b =  M-\frac{Q^2}{2r} 
-\int_r^\infty \left ( 2\pi b r'^2(T_{rr}+S)+\frac{bq^2-Q^2}{2r'^2}  \right)dr' \ 
\eeq  
where the integrals are approximated numerically.
Note that, with this scheme, the outer region $r>R_{\rm out}$ where there are no particles, and hence the above integrands vanish, will be exactly \rn{} (with no truncation error), i.e. 
\begin{subequations}
\begin{align}
q(r>R_{\rm out})=Q \ , \\
b(r>R_{\rm out})=1 \ , \\
a(r>R_{\rm out})=1-\frac{2M}{r}+\frac{Q^2}{r^2} \ .
\end{align}
\end{subequations}

\section{Constructing static solutions}
\label{sec:static_soln}

\begin{figure}
\begin{center}
\includegraphics[width=\columnwidth,draft=false]{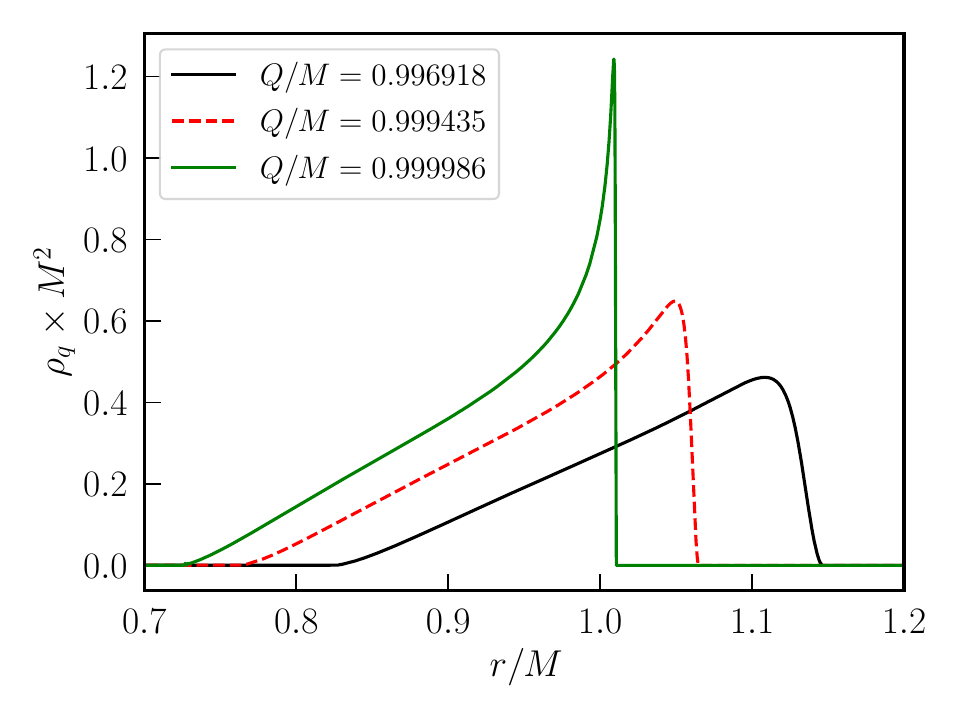}
\includegraphics[width=\columnwidth,draft=false]{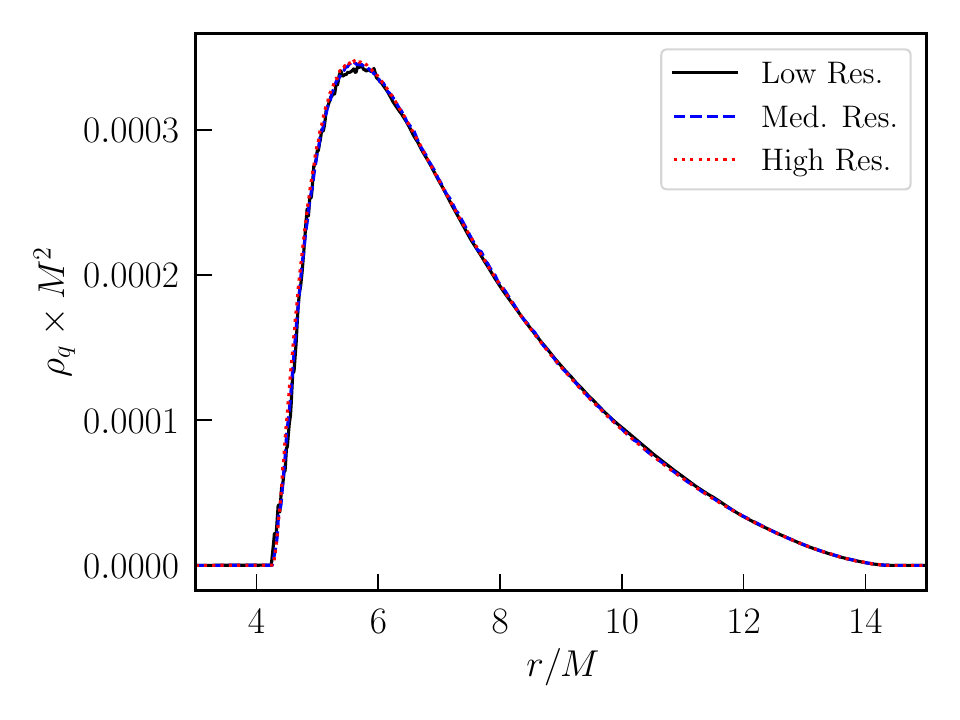}
\end{center}
\caption{
Charge density as function of areal radius.
Top: Several static shell solutions approaching $Q/M \rightarrow 1$. 
Bottom: Charge density obtained after letting the intermediate case in the top panel disperse. This is used for initial data
for the evolutions where the threshold solution is an extremal black hole. 
We show this quantity at the three different numerical resolutions described in Sec.~\ref{sec:conv}.
\label{fig:rhoq}
}
\end{figure}

Here we give more details on constructing static, spherically symmetric solutions to the Einstein-Maxwell-Vlasov 
equations following Ref.~\cite{Andreasson:2009qu}. 
Using the Killing vector $\hat{t}^a$ associated with stationarity, the conserved particle energy (per rest-mass) is
\beq
E = -\frac{1}{\mathfrak m}\hat{t}^a(p_a+\mathfrak{e}A_a) \ ,
\eeq
where $A_a$ is the vector potential satisfying 
\beq
F_{ab} = \nabla_a A_b -\nabla_b A_a \ ,
\eeq
and we will choose $\hat{t}^aA_a=0$ at $r=0$.
A distribution function that only depends on the particle energy and angular momentum automatically satisfies the static Vlasov equation.
We use a slight simplification
of the solutions constructed in Ref.~\cite{Andreasson:2009qu} and consider a distribution
function that is a linear function of energy up to some maximum value $E_0$, with a single allowed value of angular momentum
\beq
f(E,\ell) \propto
\begin{cases} 
(1-E/E_0)\delta(\ell-\ell_0) & \text{if } E<E_0 \\
0 & \text{otherwise} \ .
\end{cases}
\eeq
We solve 
for the static solution in Schwarzschild coordinates: 
\beq
ds^2=-e^{2\mu}d\tilde{t}^2+e^{2\lambda}dr^2+r^2d\Omega^2 \ . 
\eeq
After fixing $\ell_0$ and $q_0$, one obtains a family of solutions by choosing different values of $e^{2\mu}/E_0^2$ at $r=0$
and then integrating the ordinary differential equations determining the electric field and metric (see Ref.~\cite{Andreasson:2009qu}). These solutions are shell-like in that the matter is zero for $r<R_{\rm in}$
with 
\beq
R_{\rm in}= \ell_0 \left[ E_0^2e^{-2\mu}(r=0) -1 \right]^{-1/2}
\eeq
and also zero for $r>R_{\rm out}$ for finite $R_{\rm out}$, where the solution is \rn{}.

In order to evolve these solutions, we perform a coordinate transformation to outgoing co\"ordinates:
\beq
t = \tilde{t}-\int_0^r e^{\lambda-\mu} dr'
\eeq
which gives $a=e^{-2\lambda}$ and $b=-e^{\lambda+\mu}$.
To approximate the initial distribution function, we use $N=N_r \times N_p$ particles placed
at different radii $\{r_i\}$ with $i=1,\ldots,N_r$ defined (in terms of the floor function) by 
\beq
i = \lfloor N_rq(r_i)/Q+1/2 \rfloor \ ,
\eeq 
where at each position there are $N_p$ particles with different values of the radial momentum $\{p_r^j\}$ given by 
\beq
j = \lfloor N_p P(r_i,p_r^j)/P(r_i,p_r^{\rm max}) +1/2 \rfloor \ ,
\eeq 
with
\beq
P(r,p_r) := \int_{p_r^{\rm{min}}}^{p_r}\left [1-E(r,p_r')/E_0\right ]dp_r' .
\eeq 
The limits $p_r^{\rm min}$ and $p_r^{\rm max}$ are, respectively, the minimum and maximum values compatible with $E=E_0$.

In the top panel of Fig.~\ref{fig:rhoq}, we show the charge density for several of the static shell solutions
with $q_0=1.25$
approaching $Q/M \rightarrow 1$. The intermediate value of $Q/M$ shown there is the highest value that we use when
following the dispersal of one of these static solutions. As described above, we evolve the static solutions until
they sufficiently disperse, and then adjust $\ell$ and $q_0$ and use the time reversal as initial data. 
An example of the charge density profile from initial data constructed this way is shown in the bottom panel of Fig.~\ref{fig:rhoq}.

\section{Convergence results}
\label{sec:conv}

\begin{figure}[htb!]
\begin{center}
\includegraphics[width=\columnwidth,draft=false]{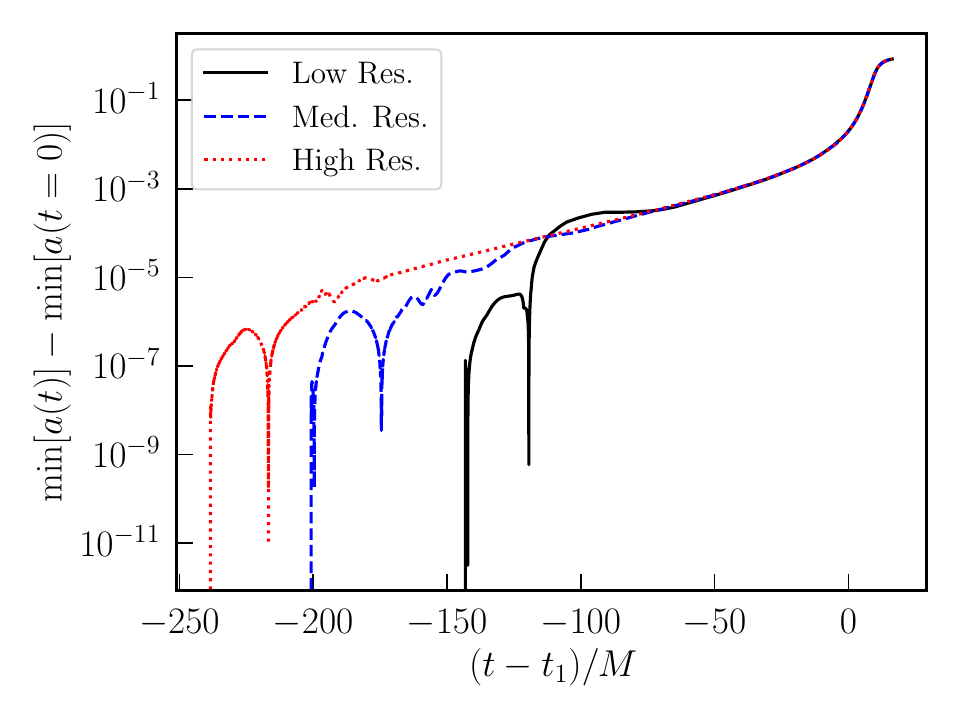}
\end{center}
\caption{
Resolution study of an unstable static solution (with $Q/M\approx 0.9994$) dispersing when evolved in outgoing coordinates. 
We show the difference in the minimum value of the metric function $a$ from its initial value.   
The times have been shifted to align the curves when $\min(a)=0.02$
\label{fig:conv_static}
}
\end{figure}

\begin{figure}[htb!]
\begin{center}
\includegraphics[width=\columnwidth,draft=false]{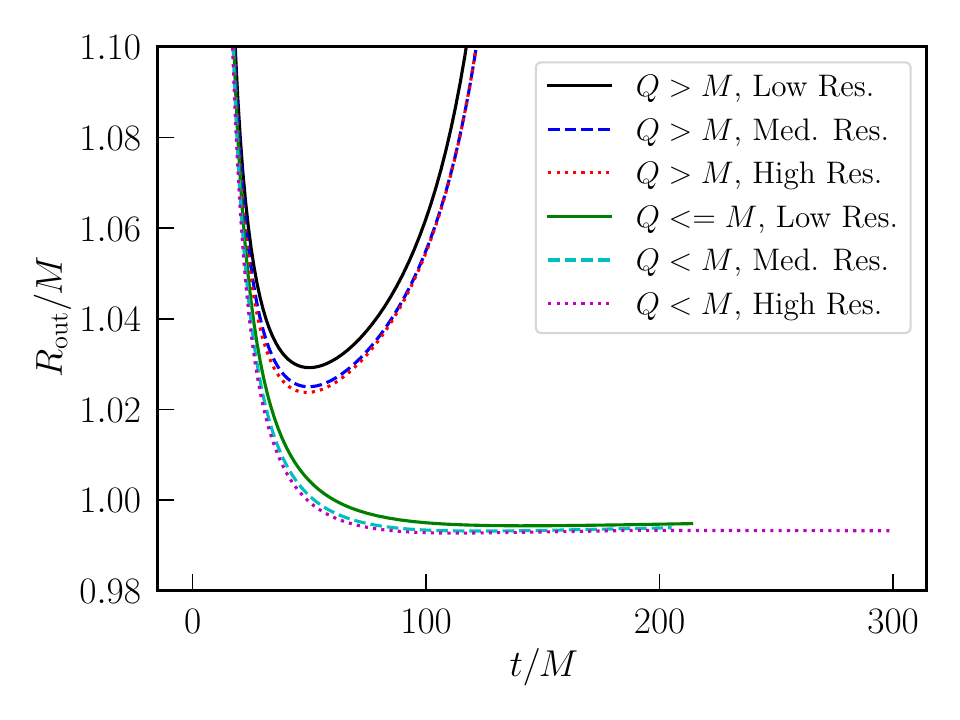}
\includegraphics[width=\columnwidth,draft=false]{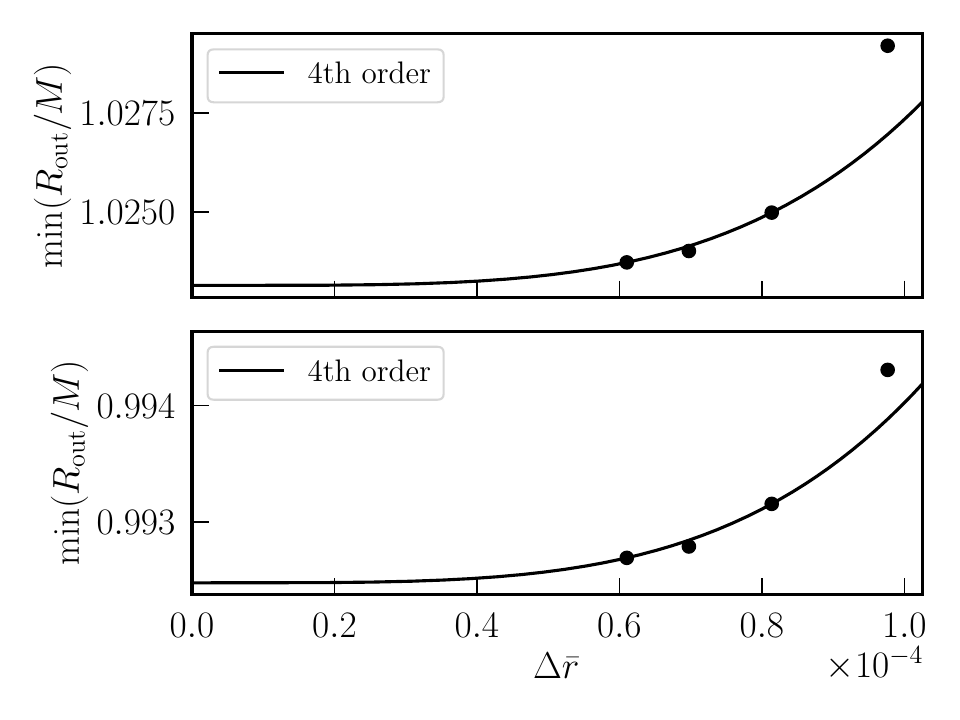}
\end{center}
\caption{
Top: Resolution study of two cases, above and below the threshold for the formation of an extremal
black hole, each at three different numerical resolutions. We show the outer radius, beyond which the solution is electrovacuum, as a function of
advanced time, around when it reaches a minimum. 
Bottom: The minimum value of the outer radius as a function of the grid spacing above (lower subplot) and below (upper subplot) the threshold
for black hole formation. 
In addition to the resolutions shown in the top panel, we show an additional one intermediate to the medium and high resolutions. 
The black curve indicates the expected trend for fourth order convergence using the medium and high resolutions.
\label{fig:conv}
}
\end{figure}

The numerical grid we use is uniformly spaced in the compactified co\"ordinate $\bar{r}\in[0,1]$. For
most of the results we present in this study, we use a grid spacing of $\Delta \bar{r}=1/12288$, along with $N_r=18432$
and $N_p=1536$ for a total of $N\approx2.8\times10^7$ particles.
For select cases, we repeat the computation with lower and higher resolution where the grid spacing is, respectively, 
$1.2 \Delta \bar{r}$ and $0.75 \Delta \bar{r}$,
and we use $(5/6)^3N$ and $(4/3)^3N$ particles.
In Fig.~\ref{fig:conv_static}, we show a resolution study of a static solution near the critical point dispersing, due to just
truncation error exciting an exponentially growing mode. As expected, as the resolution is increased, the unstable mode is seeded at a smaller and smaller amplitude.
In the top panel of Fig.~\ref{fig:conv}, 
we show a comparison of the outer radius of the particles as a function of time 
for incoming shell cases above, and just below the threshold for formation of an extremal black hole, each at the three different resolutions.
(For the case where a black hole forms, it is essentially extremal: $1-Q/M\approx10^{-12}$.)
We note that the same resolution is used both for constructing the initial data (by following the dispersal of the static shell) as for the evolution
of the incoming shell.
We can see that finite resolution leads to an overestimate of the minimum outer radius of the particles. For $Q>M$, the difference between the medium and high resolutions in $\min (R_{\rm out})$ is $\sim0.1\%$, while for the $Q<M$ case it is $0.05\%$.
In the bottom panel of Fig.~\ref{fig:conv}, we show how this quantity varies with the grid spacing.
Though evaluating the overall truncation error is somewhat complicated by the fact that it 
depends on both the number of grid points and particles, we roughly find the convergence in this quantity to be consistent with fourth order in the grid spacing, with some indication of a steeper variation at lower resolution.
Assuming fourth order convergence, we estimate the error in $\min (R_{\rm out})$ for the highest resolution to be a factor of $\sim2$  smaller than the difference between the medium and high
 resolution.

\end{document}